# Quantum Hall effect at cleaved InSb surfaces and low-temperature annealing effect

Ryuichi Masutomi,* Masayuki Hio, Toshimitsu Mochizuki, and Tohru Okamoto
*Department of Physics, University of Tokyo, 7-3-1 Hongo, Bunkyo-ku, Tokyo 113-0033, Japan*
(Dated: April 6, 2007)

We have performed low-temperature in-plane magnetotransport measurements on two-dimensional electron systems induced by deposition of Ag at *in situ* cleaved surfaces of *p*-type InSb. The quantum Hall effect was observed even at low magnetic fields around 2 T. The surface electron density and the electron mobility exhibit strong dependence on the Ag-coverage and the annealing temperature in the range of 15-40 K. The annealing effect suggests that the surface morphology strongly affects the properties of the two-dimensional electron systems.

A two-dimensional electron system (2DES) at low temperatures and in a strong magnetic field shows the quantum Hall (QH) effect, in which the diagonal resistivity $\rho_{xx}$ vanishes and the Hall resistance is quantized as $R_H = h/ie^2$ for an integer $i$ [1, 2]. Recently, Tsuji *et al.* performed in-plane magnetotransport measurements on 2DESs induced by deposition of Ag atoms at *in situ* cleaved surfaces of *p*-type InAs and observed the QH effect for $i = 4$ [3, 4]. However, a high magnetic field above 10 T was required for the observation of the QH effect. For future research on microscopic current distribution, such as edge channel transport, surface QH systems that can be observed at low magnetic fields are desirable to reduce the difficulty of constructing an experimental system with a scanning microscope. The integer QH effect is expected to appear in the region where the Landau level separation $\hbar\omega_c$ exceeds the level broadening $\hbar\tau^{-1}$ ($\omega_c\tau > 1$). Here, $\omega_c$ is the cyclotron frequency and $\tau$ is the scattering time. The dimensionless parameter $\omega_c\tau$ is equivalent to the product of $B$ and the electron mobility $\mu$. It is well-known that InSb has a very small electron effective mass $m^* = 0.013 m_e$ at the $\Gamma$ point, where $m_e$ is the free-electron mass. This value is about a half of that in InAs ($m^* = 0.026 m_e$) and thus higher electron mobility ($\mu = e\tau/m^*$) is expected. In this letter, we report the observation of the QH effect at cleaved surfaces of InSb. The quantized Hall resistance and the vanishing of $\rho_{xx}$ were clearly observed even at low magnetic fields around 2 T. Besides the dependence on the coverage of adsorbed Ag atoms, annealing effect was studied in the range of 15-40 K. The results suggest that the surface morphology strongly affects the properties of 2DESs.

The samples used were cut from a Ge-doped single crystal with an acceptor concentration of $1\text{-}2 \times 10^{21}$ m$^{-3}$. In the case of *p*-type crystals, the surface electrons are separated from the three-dimensional carriers in the substrate by the depletion layer. Sample preparation and experimental setup are similar to those used in the previous work on InAs surfaces [3, 4]. Sample cleaving was done at low temperatures in an ultra-high vacuum chamber with a $^4$He cryostat. Magnetotranport data on the (110) cleaved surface were taken in a Hall bar geometry (4 mm × 0.4 mm) using the standard four-probe lock-in technique at 13 Hz with two current electrodes and four voltage electrodes prepared by deposition of gold films onto noncleaved surfaces at room temperature. Before Ag deposition on the cleaved surface, resistance between any two electrodes was greater than 1 MΩ at 2.0 K. A magnetic field $B$ was applied in the perpendicular direction with respect to the cleaved surface. All the data shown here were taken on optically flat surfaces at 2.0 K.

In Fig. 1(a), the electron densities and mobilities are shown as a function of Ag-coverage $\Theta$. One monolayer (ML) is defined as Ag atomic densities equivalent to the surface atomic density of InSb(110) ($6.74 \times 10^{18}$ atoms/m$^2$). Due to low density of states in InSb 2DESs, more than one subband is occupied even at moderate total electron densities [5, 6]. The electron density $N_0$ ($N_1$) of the lowest (the first excited) subband was deduced from the analysis of the Shubnikov-de Hass (SdH) oscillations, while the total electron density $N_s$ was obtained from the Hall coefficient in low magnetic fields. The $N_s$-dependence of the subband occupation almost agrees with that reported by Därr *et al.* [6], Takada *et al.* [7], and Ando [8]. In the $\Theta$-region studied, $N_s$ gradually decreases with increasing $\Theta$. The negative $\Theta$-dependence of $N_s$ was also observed at InAs cleaved surfaces for Ag-coverage of $\Theta \geq 0.04$ ML [3, 4]. It is consistent with the negative coverage dependence of the Fermi energy observed in photoelectron and electron-energy-loss spectroscopy measurements on cleaved surfaces of InAs and InSb with various kinds of adsorbed materials [9–21]. The Fermi energy $\varepsilon_F$ is considered to be at an adsorbate-induced surface donor level located in the conduction band except in the very low-$\Theta$ region where the donors are fully ionized [3, 4, 22]. The observed results indicate the reduction of the energy of valence electron of donors with increasing $\Theta$. It can be explained by overlapping of the wave functions of neighboring adsorbates.

The transport mobility $\mu_t$ obtained from the zero-field conductivity $\sigma_0$ ($\mu_t = \sigma_0/eN_s$) is one order of magnitude higher than that measured for InAs surfaces [3, 4]. This may be attributed to lower acceptor concentration of the substrate as well as to smaller electron effective mass [23]. Although the origin of the strong positive $\Theta$-dependence of $\mu_t$ is not clear at the present stage, it is considered to be related to the arrangement of the adsorbates on the surface. From the analysis of the SdH oscillations

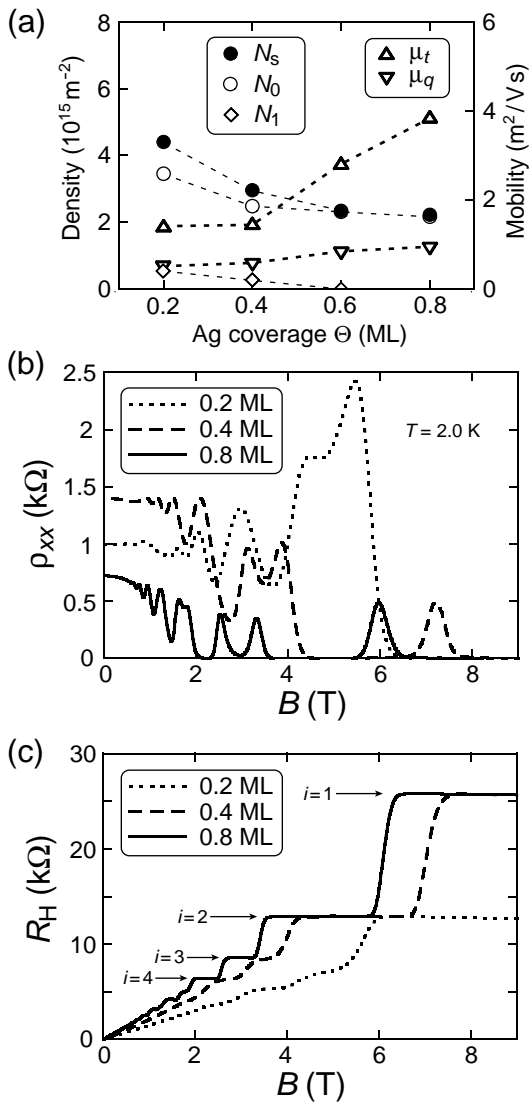

FIG. 1: (a) Electron densities (left axis) and electron mobilities (right axis) as a function of Ag coverage. Dotted lines are guides to the eye. (b) Diagonal resistivity and (c) Hall resistance at 2.0 K as a function of $B$ for different $\Theta$.

[24], the quantum mobility $\mu_q$ was deduced [25, 26]. It is known that $\mu_q$ and $\mu_t$ are identical for short-range potential fluctuations but are different for long-range potential fluctuations [25, 26]. The obtained $\mu_q$, which is much smaller than $\mu_t$, suggests that long-range potential fluctuations suppress the quantum lifetime $\tau_q$.

In Figs. 1(b) and 1(c), $\rho_{xx}$ and $R_H$ are shown as a function of $B$ for different $\Theta$, respectively. The SdH oscillations and integer QH effect are observed for all $\Theta$. The vanishing of $\rho_{xx}$, which implies zero diagonal conductivity, demonstrates that parallel conduction via the submonolayer adsorbates is negligible. For $\Theta = 0.2$ and 0.4 ML, both the lowest and first excited subbands form the Landau level structures and contribute to the magnetoconductance. The QH effect occurs only when $\varepsilon_F$ is in the localized states between the Landau levels for both subbands. For $\Theta = 0.8$ ML, on the other hand, only the lowest subband is occupied. Furthermore $\mu_t$ and $\mu_q$ reach their highest values. The QH effect with $i = 4$ is observed at 2.3 T with $\omega_c \tau_q = \mu_q B = 2.1$. The magnetic field is much lower than 10.8 T for the previous observation at an InAs surface [3, 4]. The odd-integer QH states with $i = 1$ and 3, which correspond to the spin splitting, are also observed. The QH effect obtained at a rather low magnetic field in a $^4$He cryostat will lead to further studies using scanning probe microscopy techniques.

Since the deposition of Ag was done at a low cryostat temperature below 3 K, the surface morphology is expected to change after raising the temperature. Annealing effect was studied for $\Theta = 0.8$ ML for which the total electron density is less than $2.8 \times 10^{15}$ m$^{-2}$ and the SdH oscillations arising from excited subbands are not observed. A thermal cycle between the measurement temperature of 2.0 K and a desired temperature $T_a$ with a period of about 10 min was repeated. The annealing temperature $T_a$ was increased from 10 K to 40 K with a step of 5 K.

In Fig. 2(a), $N_s$, $\mu_t$ and $\mu_q$ are shown as a function of $T_a$. While annealing effect was small for $T_a = 10$ K, $N_s$ decreases drastically with increasing $T_a$ above 15 K. This indicates that the spatial distribution of adsorbed Ag atoms changes with $T_a$. We think that cluster formation leads to the reduction of the surface donor level. The transport mobility $\mu_t$ shows the maximum at $T_a = 25$ K and decreases for higher $T_a$ while the change in $\mu_q$ is small. The negative $T_a$-dependence of $\mu_t$ may be attributed to negative $N_s$-dependence of the scattering rate $\tau_t^{-1}$ due to screened Coulomb potential [27]. The reduction of the ratio $\mu_t/\mu_q$ with decreasing $N_s$ was also observed in a back-gated GaAs/AlGaAs modulation-doped heterostructure [28].

Figure 2(b) shows the evolution of the magnetoresistance with increasing $T_a$. The observation of the clear SdH oscillations demonstrates that the annealing does not cause macroscopic inhomogeneity of the sample for all $T_a$. Thus we can control $N_s$ using the annealing effect after the deposition of adsorbates. For $T_a = 25$ K, the vanishing of $\rho_{xx}$ is observed even at 1.6 T with $i = 4$. For $T_a \geq 30$ K, $\rho_{xx}$ increases steeply in the higher-$B$ side of the QH state with $i = 1$. In this region, $\varepsilon_F$ lies in the localized states below the center of the lowest Landau level. It seems that the mobility of the present 2DES is not high enough to observe the fractional QH effect. The optimization of the surface morphology will reduce the random potential and improve the electron mobility.

In summary, in-plane magnetotransport measurements have been performed on adsorbate-induced 2DESs formed at *in situ* cleaved surfaces of $p$-type InSb. The quantum Hall effect was observed at much lower magnetic fields than that for the observation at an InAs surface due to higher electron mobility. We found that low-temperature annealing drastically changes $N_s$ and $\mu_t$. This suggests that the surface morphology strongly affects the properties of 2DESs. We hope that this work will stimulate further studies combined with scanning probe microscopy techniques.

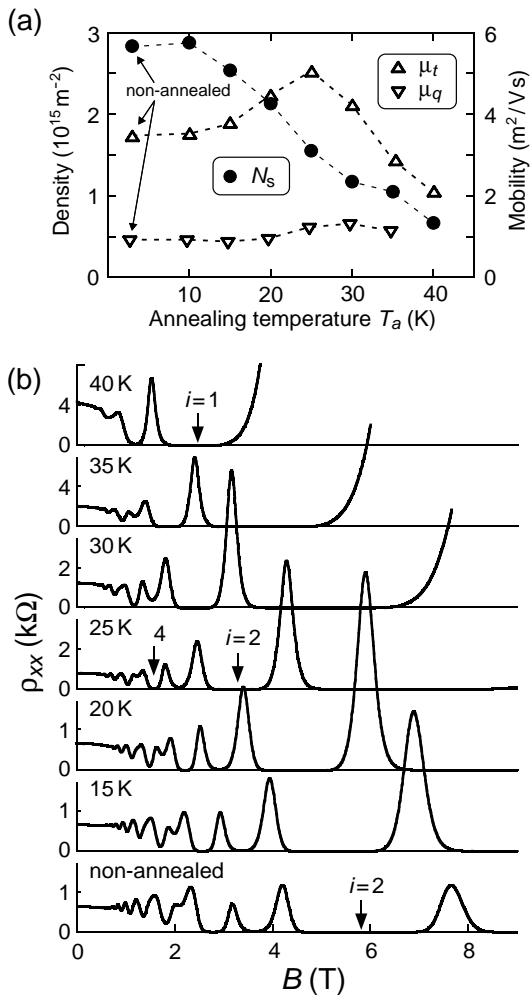

FIG. 2: (a) Electron density and mobilities as a function of annealing temperature. Dotted lines are guides to the eye. (b) Evolution of the $B$-dependence of $\rho_{xx}$ with increasing annealing temperature (from bottom to top).

This work was supported by Sumitomo Foundation, Grant-in-Aid for Scientific Research (B) (No. 18340080), Grant-in-Aid for Scientific Research on Priority Area "Physics of new quantum phases in superclean materials" (No. 18043008), Grant-in-Aid for Young Scientists (B) (No. 17740218) and Grant-in-Aid for JSPS Fellows (No. 1811418) from MEXT, Japan.